\documentclass[journal]{IEEEtran}
\usepackage{graphicx}
\usepackage{subcaption}
\usepackage{epsfig}
\usepackage{amsmath}
\usepackage{amssymb}
\usepackage{algorithm,algorithmic}
\usepackage{amsmath,amsfonts,amssymb}
\usepackage{mathtools}
\usepackage{cite}
\usepackage[space]{grffile}
\usepackage[hyphens]{url}
\usepackage{hyperref}
\usepackage{url}
\usepackage{array}
\usepackage{fixltx2e}
\DeclareMathOperator*{\argmin}{arg\,min}
\DeclarePairedDelimiter{\norm}{\lVert}{\rVert}

\begin{document}

\title{SPAM: Signal Processing for Analyzing Malware}

\author{Lakshmanan~Nataraj,~\IEEEmembership{Student Member,~IEEE,}
        B.S.~Manjunath,~\IEEEmembership{Fellow,~IEEE}
}


\maketitle

%
\IEEEpeerreviewmaketitle




\IEEEPARstart{C}{yber} attacks have risen in recent times.
The attack on Sony Pictures by hackers, allegedly from North Korea, has caught worldwide attention. 
The President of the United States of America issued a statement and ``vowed a US response after North Korea's alleged cyber-attack".
This dangerous malware termed ``wiper" could overwrite data and stop important execution processes.
An analysis by the FBI showed distinct similarities between this attack and the code used to attack South Korea in 2013, thus confirming that hackers re-use code from already existing malware to create new variants.
This attack along with other recently discovered attacks such as Regin, Opcleaver give one clear message: \emph{current cyber security defense mechanisms are not sufficient enough to thwart these sophisticated attacks.}

Today's defense mechanisms are based on scanning systems for suspicious or malicious activity.
If such an activity is found, the files under suspect are either quarantined or the vulnerable system is patched with an update. 
These scanning methods are based on a variety of techniques such as static analysis, dynamic analysis and other heuristics based techniques, which are often slow to react to new attacks and threats.
\emph{Static analysis} is based on analyzing an executable without executing it, while \emph{dynamic analysis} executes the binary and studies its behavioral characteristics.
Hackers are familiar with these standard methods and come up with ways to evade the current defense mechanisms.
They produce new \emph{malware variants} that easily evade the detection methods.
These variants are created from existing malware using inexpensive easily available ``factory toolkits" in a ``virtual factory" like setting, which then spread over and infect more systems.
Once a system is compromised, it either quickly looses control and/or the infection spreads to other networked systems.  
While security techniques constantly evolve to keep up with new attacks, hackers too change their ways and continue to evade defense mechanisms.
As this never-ending billion dollar ``cat and mouse game" continues, it may be useful to look at avenues that can bring in novel alternative and/or orthogonal defense approaches to counter the ongoing threats.
The hope is to catch these new attacks using orthogonal and complementary methods which may not be well known to hackers, thus making it more difficult and/or expensive for them to evade all detection schemes.
\emph{This paper focuses on such orthogonal approaches from Signal and Image Processing that complement standard approaches.}

\section*{MALWARE LANDSCAPE}

\begin{figure}[t]
\centering
{\includegraphics[width=\linewidth, height=0.6\columnwidth]{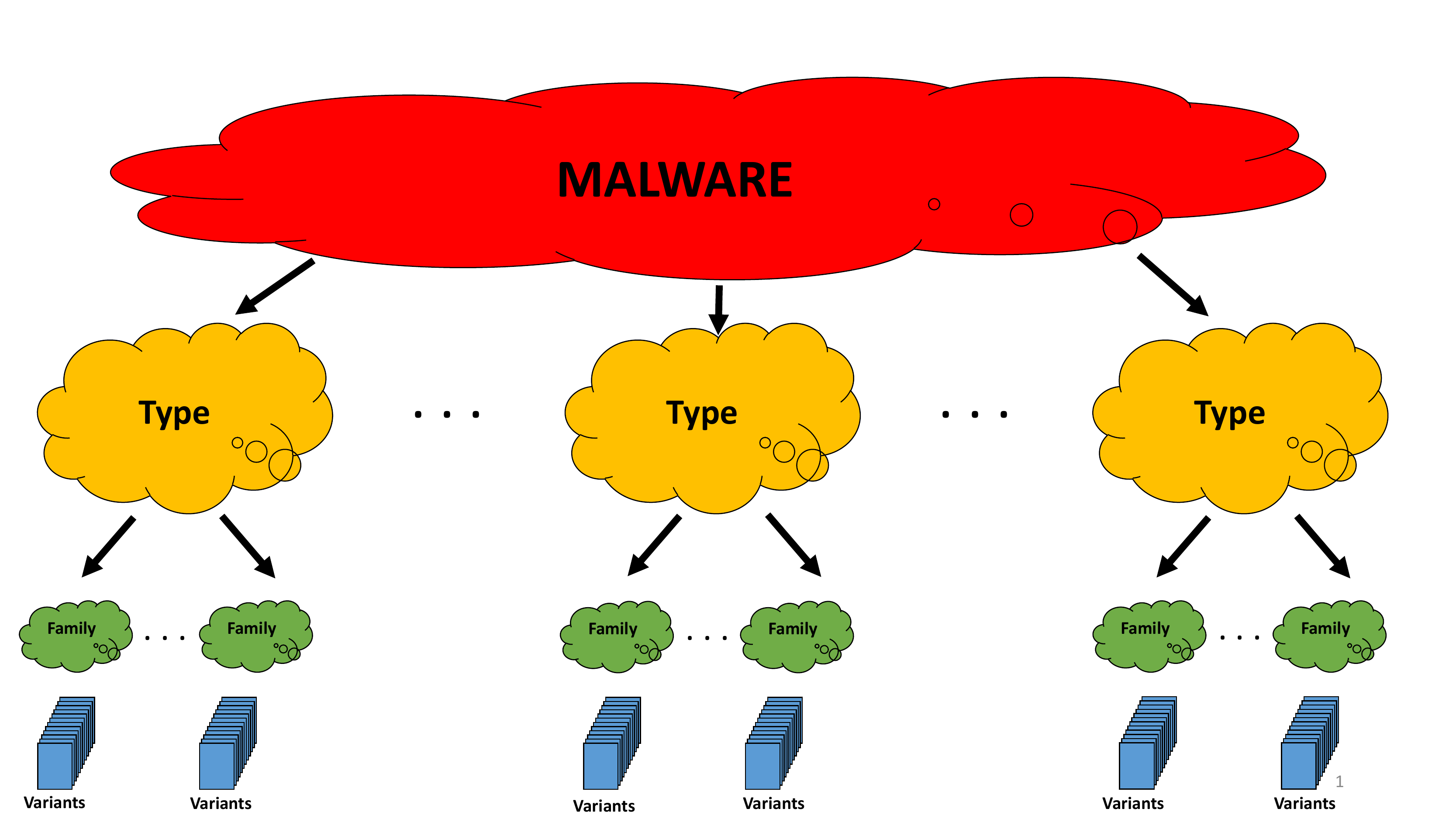}}
\vspace{-10pt}
\caption{Malware Landscape}
\label{fig-landscape}
\vspace{-20pt}
\end{figure}

 Malware - malicious software, is any software that is designed to cause damage to a computer, server, network, mobile phones and more such devices.
Based on their function, malware are classified into different \emph{Types} such as Trojans, Backdoors, Virus, Worm, Spyware, Adware and more.
Malware are also identified by which \emph{Platform} they belong to, such as Windows, Linux, AndroidOS and others.
Apart from \emph{Types} and \emph{Platforms}, malware are further classified into \emph{Families} depending on their specific function. 
These \emph{Families} in turn have many \emph{Variants} which perform almost the same function. 
The entire malware landscape is shown in Fig.~\ref{fig-landscape}.
According to the Computer Antivirus Research Organization (CARO) convention for naming malware, a malware is represented by: \emph{Type:Platform/Family.Variant}.
For example, \emph{PWS:Win32/Zbot.gen!AF} denotes a password stealer malware of the generic Zbot family that attacks 32-bit Windows platforms.

 Malware variants are created either by making changes to the malware code or by using executable packers.
In the former case a simple mutation occurs by changing small parts of the code. 
These are referred as \emph{unpacked malware variants}.
In the latter case a more complex mutation occurs either by compressing or encrypting (usually with different keys) the main body of the code and appending a decompression/decryption routine, which during runtime decompresses/decrypts the encrypted payload.
The new variants are called \emph{packed malware variants} and they perform the same function as the original malware but their attributes would be so different that Antivirus software, which use traditional signature based detection, would not be able to detect them.
The tools used for obfuscation are called \emph{Executable Packers}, available both as freeware and commercial tools.
There are hundreds of packers that exist today which make it very easy for malware writers to create new variants.

\begin{figure*}[t]
\centering
{\includegraphics[width=\linewidth, height=\columnwidth]{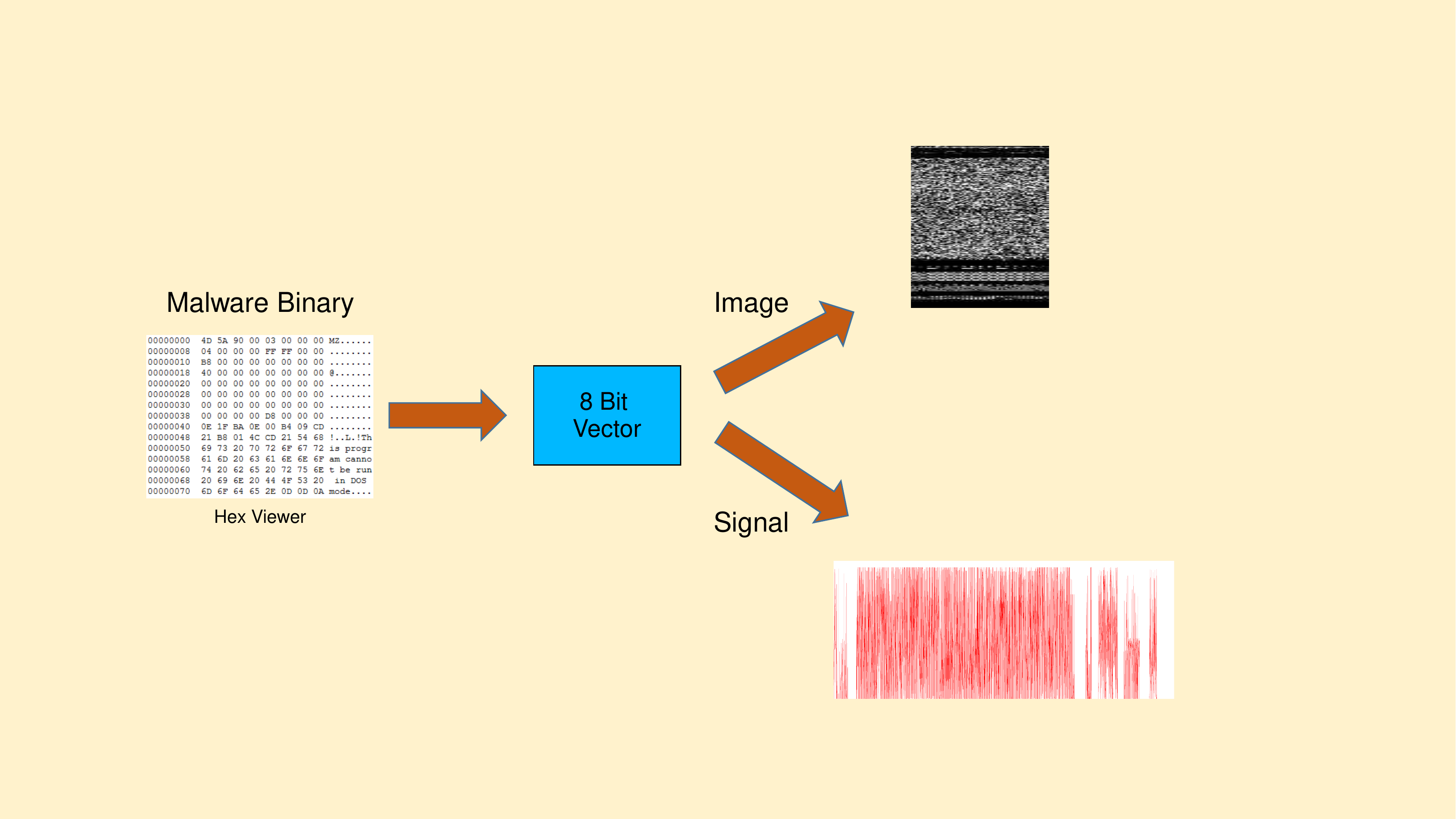}}
\vspace{-10pt}
\caption{Malware represented as a Signal and an Image}
\label{fig1}
\vspace{-10pt}
\end{figure*}

\section*{MALWARE ANALYSIS}

\emph{Malware classification} deals with identifying the family of an unknown malware variant from a malware dataset that is divided into many families. 
The level of risk of a particular malware is determined by what function it does, which is in turn reflected in its family.
Hence, identifying the malware family of an unknown malware is crucial in understanding and stopping new malware.
It is usually assumed that an unknown malware variant belongs to a known set of malware families (supervised classification).
Having a high \emph{classification accuracy} (the number of correctly classified families) is desirable.
A closely related problem is \emph{malware retrieval} where the objective is to retrieve similar malware matches for a given query from a large database of malware.
In \emph{malware detection} the problem is to determine if an unknown executable is malicious, benign or unknown. 
This problem is more challenging than malware classification where all samples are known to be malicious.
In this tutorial we will focus on \emph{malware classification} and \emph{malware retrieval}.

 While most malware are geared towards Windows Operating System, they are also quickly expanding to other avenues such as Android, Linux and OS X.
Antivirus vendor G-DATA reported that they discovered more than 1.5 Million malicious Android apps in 2014 and more than 400,000 apps in just the first quarter of 2015. 
Similarly, there has also been a stark rise in Linux malware and OS X malware. 
An important question in this context is: \emph{Can we have a single method that can detect malware irrespective of which Operating System it comes from without having to know the nuances of each system?}

 A common way to defeat static analysis is by using packers on a executable which compress and/or encrypt the executable code and create a new packed executable that mimics the previous executable in function but reveals the actual code only upon execution runtime.
Dynamic analysis is agnostic to packing but is slow and time consuming.
Further, today's malware are designed to be Virtual Machine (VM) aware, which either do not do any malicious activity in the presence of VM or attempts a ``suicide" when a VM is detected. 
The challenges here are: \emph{Can we design techniques that are fast, do not need disassembly, unpacking or execution?}

 A key emphasis in all the above-mentioned challenges is development of complementary methods that address the limitations of existing approaches.
Alternative representations of malware data such as signals or images have patterns that are not captured by standard methods.
We explore these types of representations in this paper.

\begin{figure}[h]
\centering
{\includegraphics[width=0.9\linewidth, height=0.7\columnwidth]{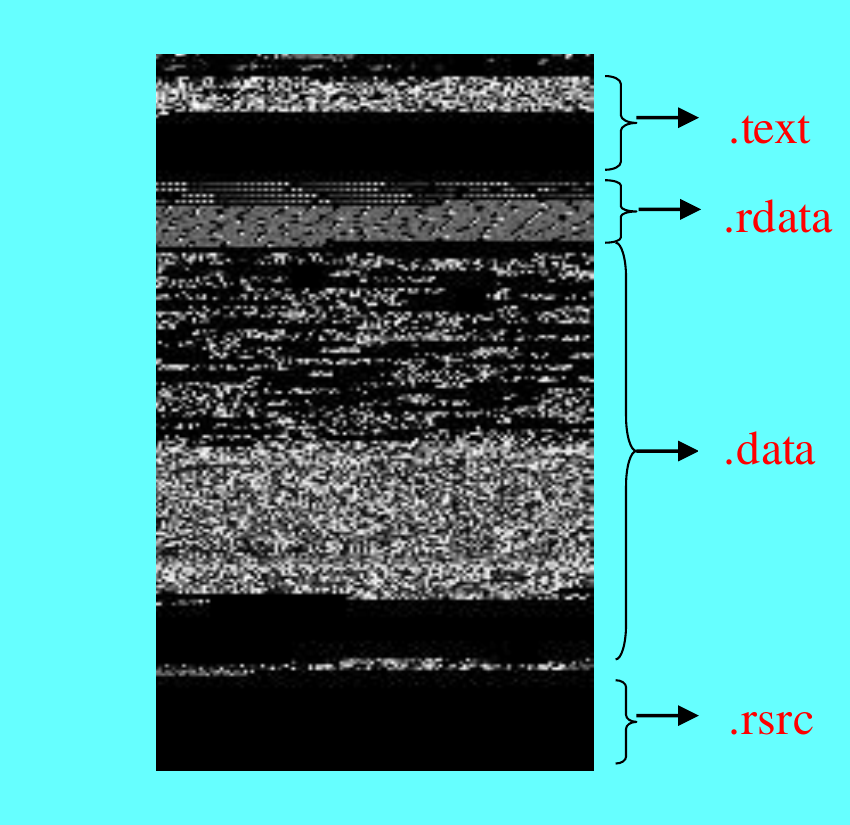}}
\caption{Example of a Malware Image}
\label{dontovo}
\vspace{-10pt}
\end{figure}

\section*{MALWARE IMAGES}

\begin{figure*}[t]
\centering
{\includegraphics[width=\linewidth, height=0.7\columnwidth]{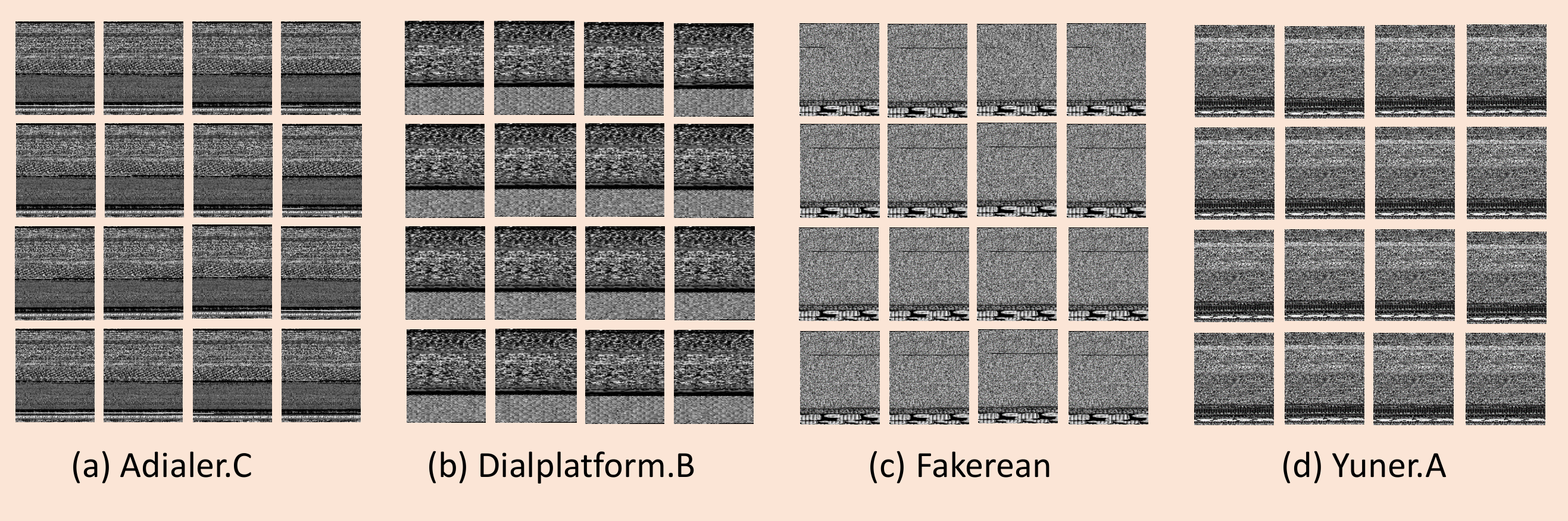}}
\vspace{-10pt}
\caption{Visual similarity among malware variants of 4 different families}
\label{fig2}
\vspace{-10pt}
\end{figure*}

 A common method of viewing and editing malware binaries is by using \emph{Hex Editors}, which display the bytes of the binaries in hexadecimal representation from `00' to `FF'.
Effectively, these are 8-bit numbers in the range of 0-255.
Grouping these 8-bit numbers results in a 8-bit vector, from which we construct a signal or an image as shown in Fig.~\ref{fig1}.
For an image, the width is fixed and the height is allowed to vary depending on the file size.
Fig.~\ref{dontovo} shows an example image of a common Windows Trojan downloader, \emph{Dontovo.A}, which downloads and executes arbitrary files.
We can see that different sections of this malware exhibit distinctive image patterns.
The \emph{.text} section which contains the executable code has a fine grained texture.
It is followed by a black block (zeros), indicating zero padding at the end of this section. 
The \emph{.data} section contains both uninitialized code (black patch) and
initialized data (fine grained texture). 
The final \emph{.rsrc} section contains all the resources of the module, including the icon of the executable. 

\begin{figure}[h]
\centering
{\includegraphics[width=\linewidth, height=0.28\columnwidth]{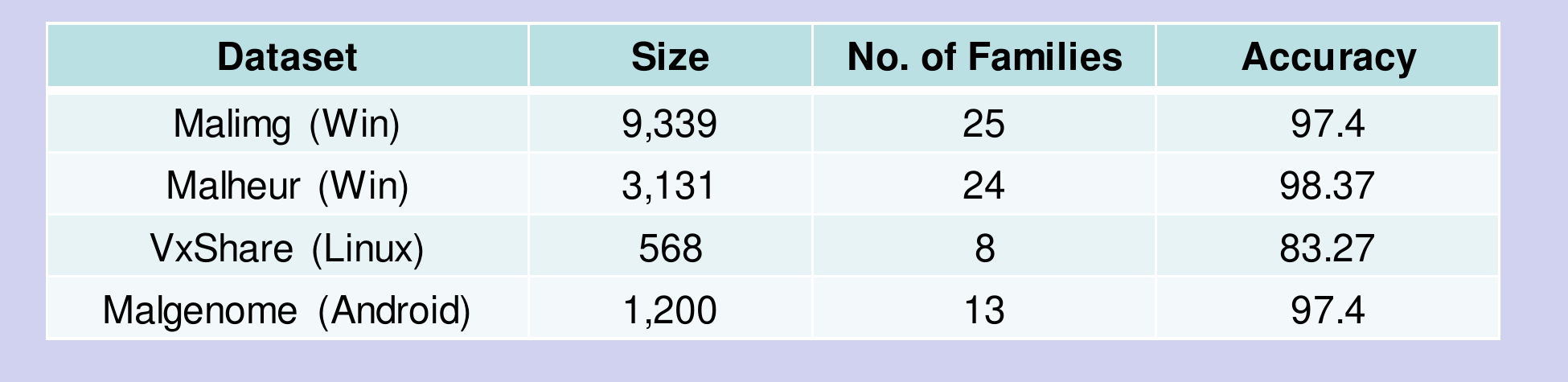}}
\caption{Classification Accuracy on Datasets from different Operating Systems}
\label{diff-os-acc}
\vspace{-10pt}
\end{figure} 

 When we visualized a large number of malware variants, an empirical observation we could make is that there was \emph{visual similarity among malware variants of the same family} (Fig.~\ref{fig2}).
At the same time, the variants were also distinct from those belonging to other families. 
This is because the variants are created using either simple code mutations or packing.
It is easy to identify the variants for unpacked malware since the structure of the variants are very similar. 
In the case of packed malware, the executable code is compressed and/or encrypted.
During runtime, this code is then unpacked and executed.
When two unpacked variants belonging to a specific malware family are using a packer to obtain packed variants of the same family, their structure no longer remains the same as that of the unpacked variants.
However, the structure within the packed variants are still similar though the actual bytes may vary due to compression and/or encryption.
The visual similarity of malware images motivated us to look at malware classification using techniques from computer vision, where image based classification has been well studied.
We use global image similarity descriptors and obtain compact signatures for these malware, which are then used to identify their families. 

\subsection*{CLASSIFICATION}
 Once the malware binary is converted to an image, an image similarity descriptor is computed on the image to characterize the malware. 
The descriptor that we use is the GIST feature~\cite{gist2}, which is commonly used in image recognition systems such as scene classification~\cite{gist2}, object recognition~\cite{gist1} and large scale image search~\cite{gist-eval}. 
Every image location is represented by the output of filters tuned to different orientations
and scales. 
A steerable pyramid with 4 scales and 8 orientations is used.
The local representation of an image is then given by: $V^{L}(x) = {V_{k}(x)}_{k=1..N}$ where $N$ = 20 is the number of sub-bands.
To capture the global image properties while retaining some local information, the mean value of the magnitude of the local features is computed and averaged over large spatial regions: $m(x) = \sum_{x'} | V(x') | W(x'-x) $ where $W(x)$ is the averaging window. 
The resulting representation is downsampled to have a spatial resolution of M$\times$M pixels (here we use M=4).
Thus the feature vector obtained is of size M$\times$M$\times$N = 320. 
For faster processing, the images are usually resized to a smaller size (we use $64 \times 64$).

 To identify malware families, we perform supervised classification with $10$-fold cross validation and compute the average classification accuracy.
We use Nearest Neighbor (NN) classifier which assigns the family of the nearest malware to an unknown malware. 
We obtained four datasets: Malimg dataset (Windows)~\cite{malimg-ds}, Malheur dataset (Windows)~\cite{malheur}, MalGenome dataset (Android)~\cite{malgenome} and VxShare ELF dataset (Linux)~\cite{vxshare}.
On all four datasets, we obtained a high classification accuracy (Fig.~\ref{diff-os-acc}).
Further, on comparing our approach with dynmaic analysis, our method was comparable in terms of classification accuracy but {\bf 4,000} times faster than dynamic analysis~\cite{comp-assm}.  
In~\cite{sigmal}, we extend our approach to separate malware from benign software.
In order to get a richer discrimination between benign and malicious samples, we adopt a \emph{section-aware} approach and compute GIST descriptors on the entire binary as well as the top two sections of the binary which could contain the code. 
With more than {\bf 99\%} precision, our approach outperformed other static similarity features.




\begin{figure*}[t]
\centering
{\includegraphics[width=0.9\linewidth, height=0.98\columnwidth]{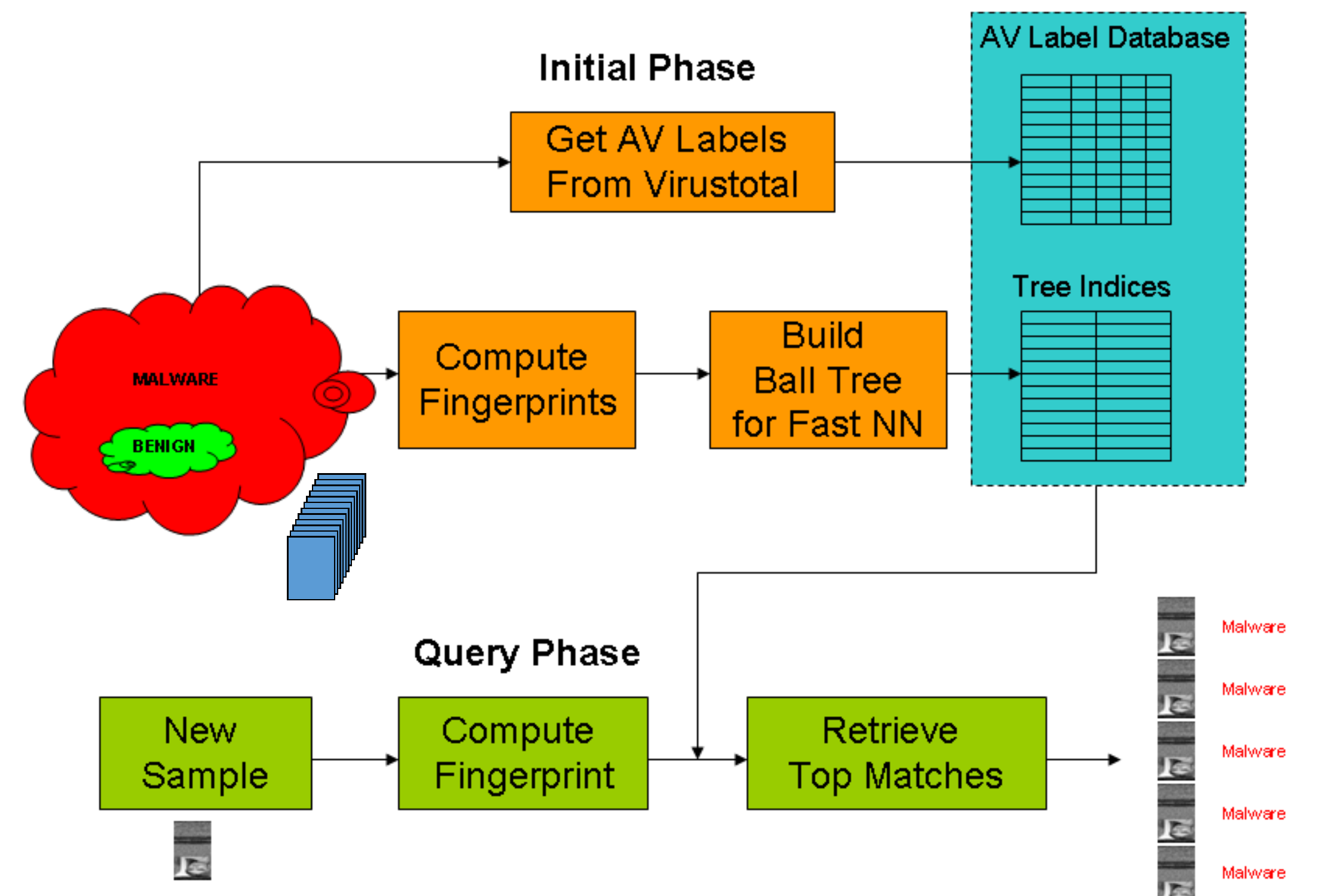}}
\caption{Block Schematic of SARVAM}
\label{fig-sarvam}
\end{figure*} 

\subsection*{SEARCH AND RETRIEVAL}

 We developed \emph{SARVAM: Search And RetrieVAl of Malware}~\cite{sarvam} (accessible at {\it http://sarvam.ece.ucsb.edu}), an online system for large scale malware search and retrieval.
It is one of the few systems available to public where researchers can upload or search for a sample and retrieve similar malware matches from a large database.
Leveraging on our past work~\cite{malw-imgs}, we use GIST descriptors for content-based search and retrieval of malware.
These effectively capture the visual (structural) similarity between similar malware variants.
For fast search and retrieval, we use a scalable \emph{Balltree}-based Nearest Neighbor searching technique.
On a database of more than {\bf 7 Million} samples comprising mostly malware and a few benign samples, SARVAM could find a match in about {\bf 6} seconds.  
SARVAM has been operational since May 2012 and during this period, we received more than 440,000 samples.
Nearly 60\% were possible variants of already existing malware from our database.

  There are two phases in the system design as illustrated in Fig.~\ref{fig-sarvam}.
During the initial phase, we first obtain a large corpus of malware samples from various sources~.
The image fingerprints for all the samples in the corpus are then computed and stored in a database.
Simultaneously, we obtain the Antivirus (AV) labels for all the samples from Virustotal~\cite{vtotal}, an online system that maintains a database of AV labels.
These labels act as a ground truth and are later used to describe the nature of a sample, i.e., how malicious or benign a sample is.
During the query phase, the fingerprint for the new sample is computed and matched with the existing fingerprints in the database to retrieve the top matches.

\begin{figure}[h]
\centering
{\includegraphics[width=3.2in,height = 1.8 in]{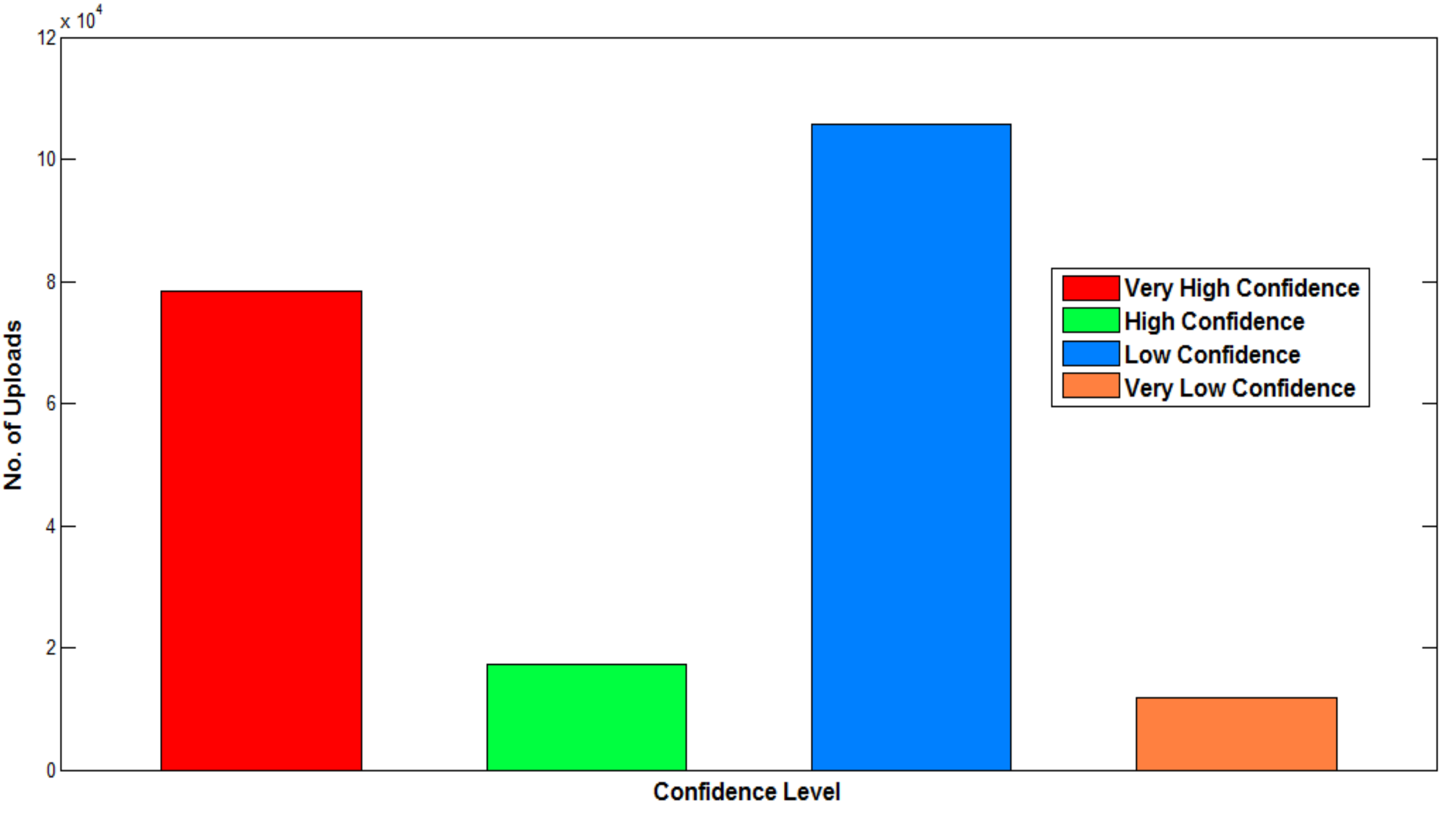}}
\caption{Confidence of the Top Match}
\label{match-conf}
\vspace{-10pt}
\end{figure}

 Of the 440,000 uploaded samples we received, not all the samples have a good match with our corpus database.
In Fig.~\ref{match-conf}, we see the distribution of the confidence levels of the top match.
Close to 37\% fall under Very High Confidence, 8\% under High Confidence, 49.5\% under Low confidence and  5.5\% under Very Low Confidence.

\section*{SPARSITY BASED MALWARE ANALYSIS}

\begin{figure*}[ht]
\centering
{\includegraphics[width=\linewidth, height=0.98\columnwidth]{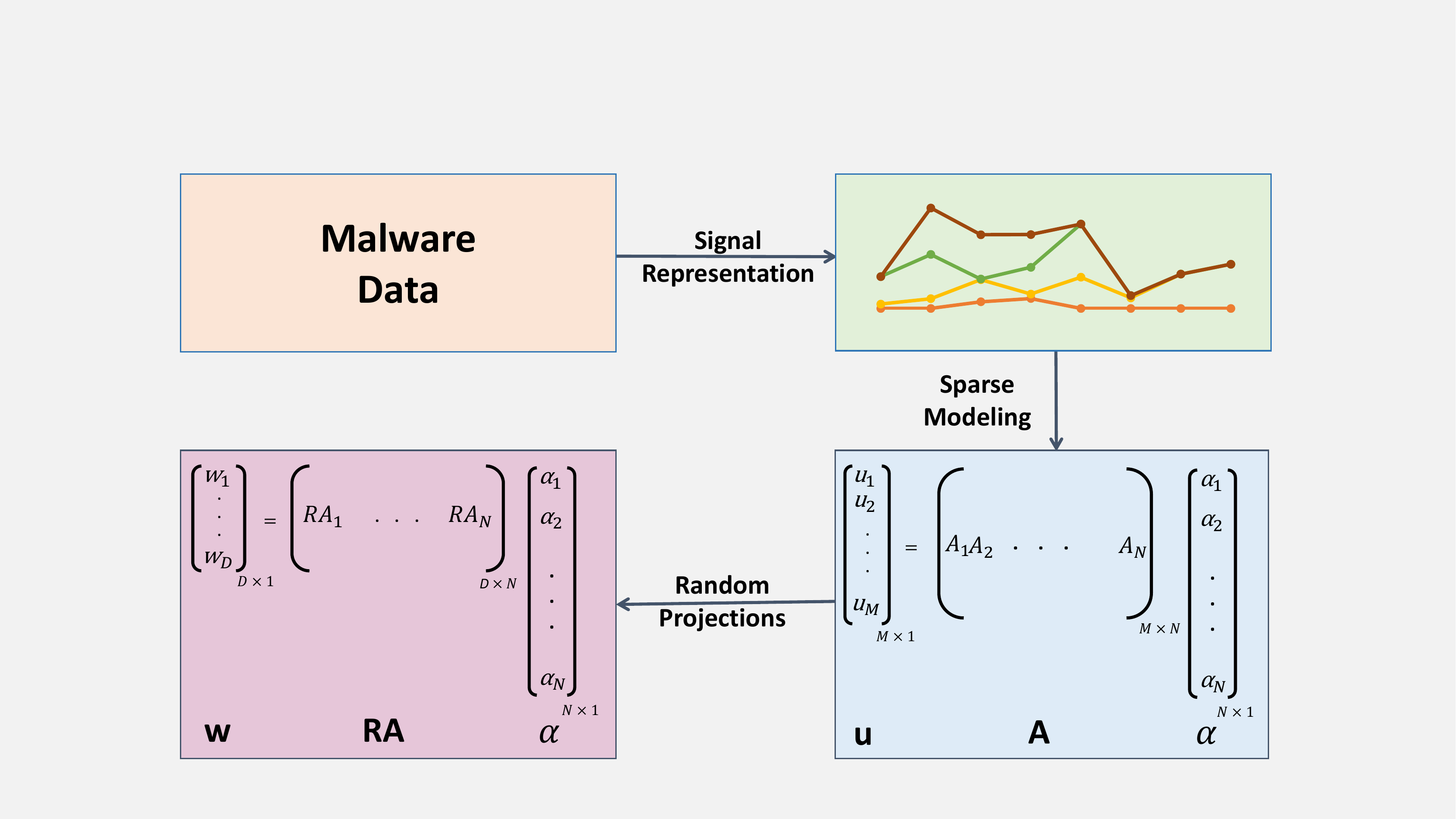}}
\vspace{-10pt}
\caption{Sparse Representation based Classification (SRC) framework for Malware Classification}
\label{fig-sp}
\vspace{-10pt}
\end{figure*}

In this part we explore Sparse Representation based Classification (SRC) methods to classify malware variants into families. 
Such methods have been previously applied to problems where samples belonging to a class have small variations in them, such as face recognition~\cite{2009-pami-wright-face} and iris recognition~\cite{2011-jay-pami}.
We developed \emph{SATTVA: SparsiTy inspired classificaTion of malware VAriants}~\cite{sattva}, where we model a malware variant belonging to a particular malware family as a linear combination of variants from that family.
Since variants of a family have small changes in the overall structure and differ from variants of other families, projections of malware in lower dimensions preserve this ``similarity''. 

Given a dataset of $N$ labeled malware belonging to $L$ different malware families with $P$ malware per family, the task is to identify the family of an unknown malware $\mathbf{u}$.
We represent a malware as a digital signal $\mathbf{x}$ of range $[0,255]$, where every entry of $\mathbf{x}$ is a byte value of the malware.  
Since each malware sample can have a different code-length, we normalize all vectors to a  maximum length ($M$) by zero-padding.

The entire dataset can now be represented as an $M \times N$ matrix $\mathbf{A}$, where every column represents a malware. Further, for every family $k$ ($k = 1,2, ... ,L$), we define an $M \times P$ matrix $\mathbf{A}_k = [\mathbf{x}_{k1},\mathbf{x}_{k2},...\mathbf{x}_{kP}]$ where $\mathbf{x}_{k\{.\}}$ represents a malware sample belonging to family $k$. Now, $\mathbf{A}$ can be expressed as a concatenation of block-matrices $\mathbf{A}_k$:
\begin{equation}
\mathbf{A} = [\mathbf{A}_1 \mathbf{A}_2 .. \mathbf{A}_L] \in {\mathbb{R}}^{M \times N}
\label{mat-form}
\end{equation} 

Let $\mathbf{u} \in \mathbb{R}^M$ be an unknown malware whose family is to be determined, with the assumption that $\mathbf{u}$ belongs to one of the families in the dataset. 
Then, following~\cite{2009-pami-wright-face}, we represent $\mathbf{u}$ as a sparse linear combination of the training samples as:
\begin{equation}\label{leq1}
\mathbf{u} = \sum\limits_{i=1}^L \sum\limits_{j=1}^P \alpha_{ij} \mathbf{x}_{ij} = \mathbf{A}\alpha
\end{equation} 
where $\alpha = [\alpha_{1,1},...,\alpha_{L,P}]^T$ represents the $N \times 1$ sparse coefficient vector ($N = LP$).
$\alpha$ will have non-zero values only for samples that are from the same family as $\mathbf{u}$. 
The sparsest solution to~\eqref{leq1} can be obtained using Basis Pursuit~\cite{2011-jay-pami} by solving the following $l_1$-norm minimization problem:
\begin{equation}
\hat{\alpha} = \argmin_{\alpha' \in \mathbb{R}^N} \ \norm{\alpha'}_1 \ \text{subject to} \ \ \mathbf{u} = \mathbf{A}\alpha'
\label{leq2}
\end{equation}

Estimating the family of $\mathbf{u}$ is done by computing residuals for every family in the training set and then selecting the family that has minimum residue.

\begin{figure*}[ht]
\centering 
\begin{subfigure}[h]{0.49\textwidth}
{\includegraphics[width=\linewidth,height=0.65\columnwidth]{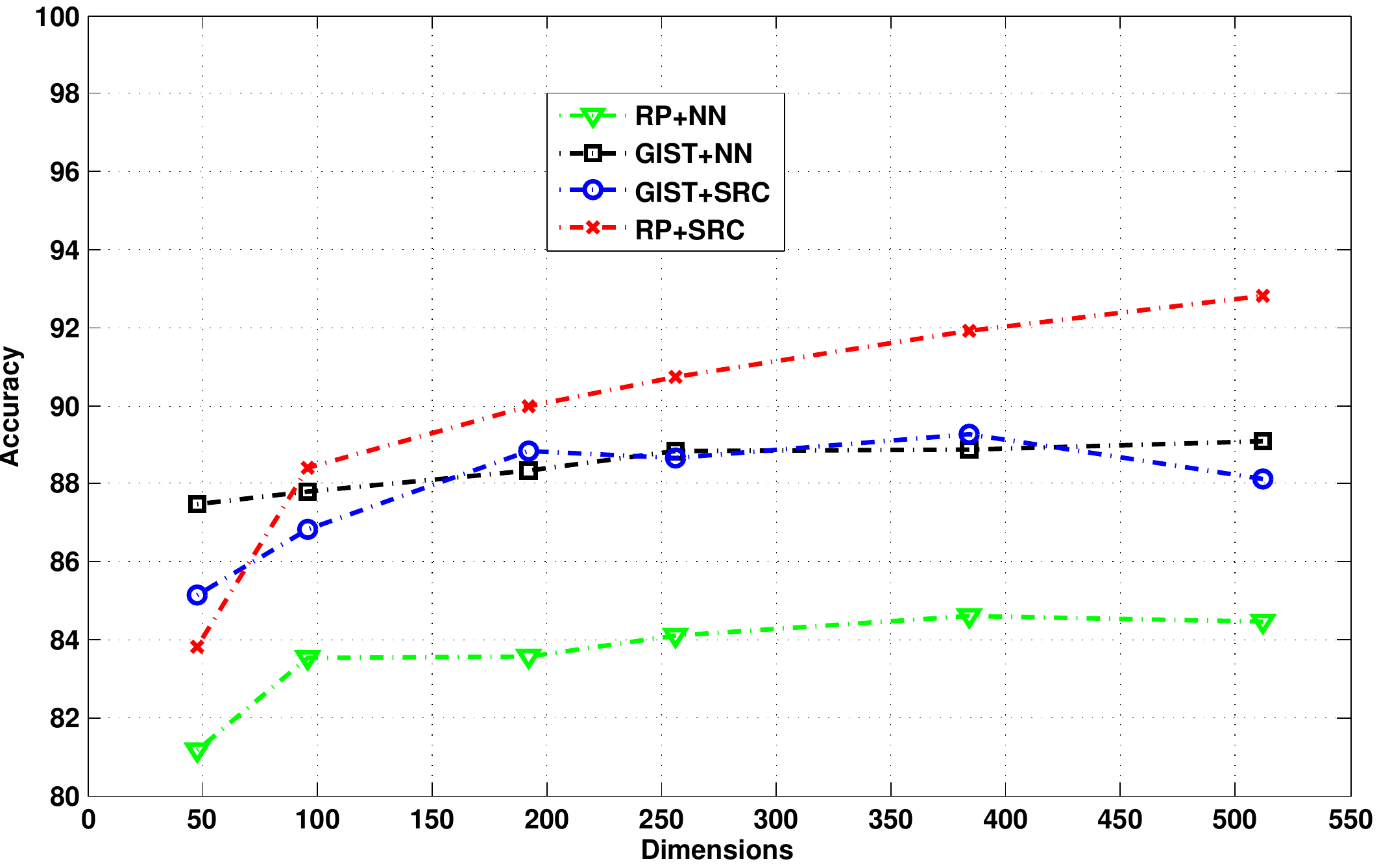}}    
\caption{Malimg Dataset}
\label{exps-res-malimg}
\end{subfigure}
\begin{subfigure}[h]{0.49\textwidth}
{\includegraphics[width=\linewidth,height=0.65\columnwidth]{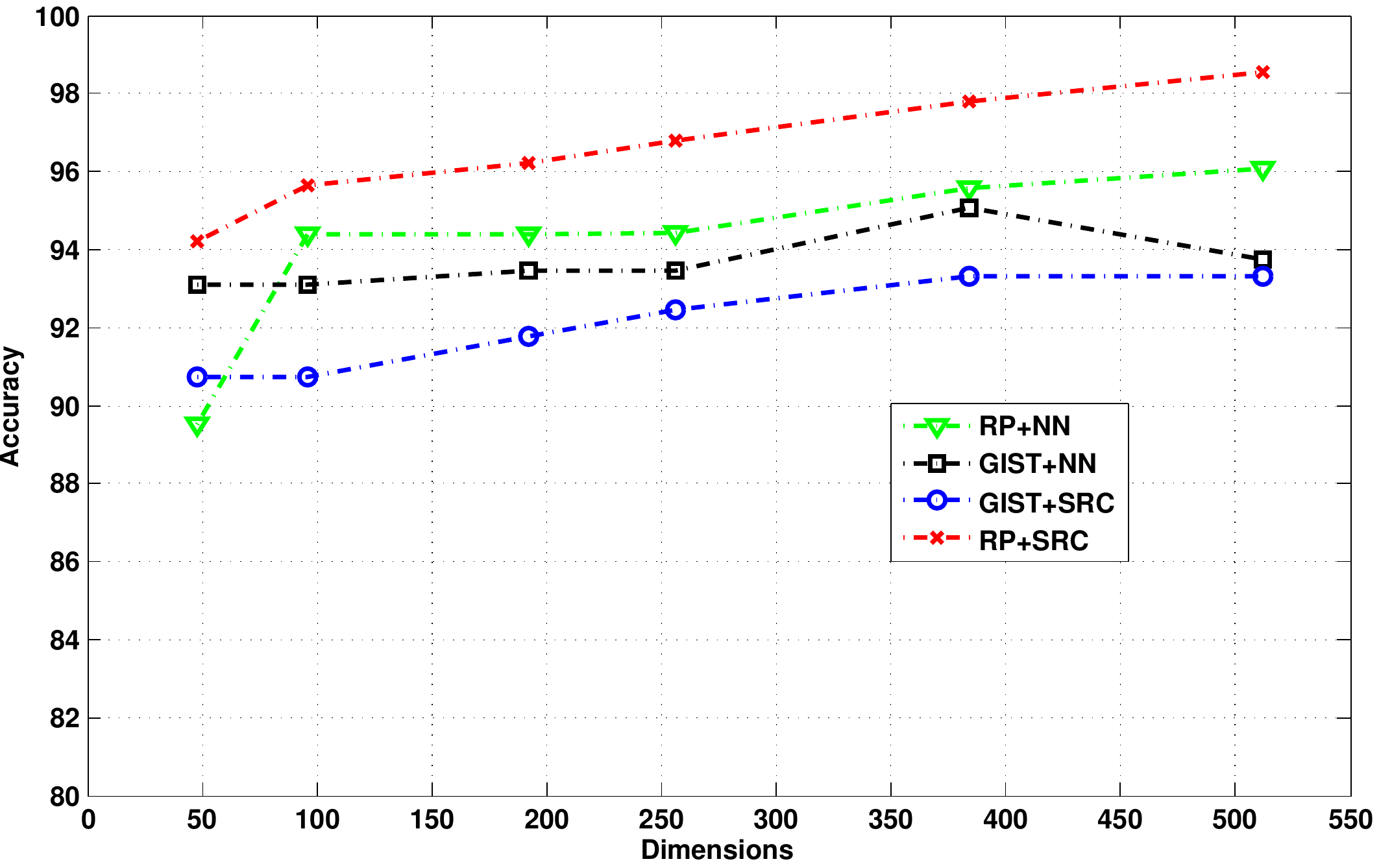}} 
\caption{Malheur Dataset}
\label{exps-res-malheur}
\end{subfigure}
\vspace{-5pt}
\caption{Experimental Results on (a) Malimg Dataset and (b)  Malheur Dataset with features using Random Projections (RP) and GIST, and classification algorithms using Sparse Representation based Classification (SRC) and Nearest Neighbor (NN).
}
\label{exps-res}
\vspace{-5pt}
\end{figure*}

\subsection*{RANDOM PROJECTIONS}

When a malware binary is represented as a numerical vector by considering every byte, the dimensions of that vector can be very high.
For example, a 1 MB malware has around 1 Million bytes and this could make the calculations computationally expensive.
Hence, we project the vectors to lower dimensions using Random Projections (RP). 
This also removes dependency on any particular feature extraction method.
Previous works have demonstrated that SRC is effective in lower-dimensional random projections as well, see~\cite{2009-donoho-counting,2009-pami-wright-face, 2011-jay-pami}.
Let $\mathbf{R} \in {\mathbb{R}}^{D \times M}$ be the matrix that projects $\mathbf{u}$ from signal space $M$ to $\mathbf{w}$ of lower dimensional space $D$ ($D<<M$):

\begin{equation}
\mathbf{w} = \mathbf{R}\mathbf{u} = \mathbf{R}\mathbf{A}\alpha
\end{equation} 
The entries of $\mathbf{R}$ are drawn from a zero mean normal distribution.
The above system of equations is underdetermined and sparse solutions can be obtained by reduced $l_1$-norm minimization:

\begin{equation}
\hat{\alpha} = \argmin_{\alpha' \in \mathbb{R}^N} \ \norm{\alpha'}_1 \ \text{subject to} \ \ \mathbf{w} = \mathbf{R}\mathbf{A}\alpha' 
\label{src-rp}
\end{equation}
The overall approach is shown in~Fig.\ref{fig-sp}.

We tested our technique on two public malware datasets: Malimg Dataset~\cite{malimg-ds} and Malheur Dataset~\cite{malheur}.
On both datasets, we selected equal number of samples to reduce any bias towards a particular family.
For comparison, we used GIST descriptors, which we had previously applied for malware classification.
We used the SRC framework to identify the malware family of a test sample and compared with Nearest Neighbors (NN) classification that was previously used in~\cite{malw-imgs}.
We varied the dimensions from $\{48,96,192,256,384,512\}$, which are consistent for both RP and GIST.
In our experiments, we chose 80\% of a dataset for training and 20\% for testing. 
On both the Malimg dataset (Fig.~\ref{exps-res-malimg}) and the Malheur dataset (Fig.~\ref{exps-res-malheur}), the best accuracy is obtained for the combination of Random Projections (RP) and the SRC classification framework.
The accuracies for GIST for both classifiers were almost the same.
In~\cite{sattva}, we further showed how this approach can be used to reject potential outliers in a dataset and also evaluated on large scale datasets having 42,480 malware and 2,124 families.


\section*{FUTURE DIRECTIONS}
While we explored signal and image based analysis of malware data, a natural complement is to treat the malware as audio-like one dimensional signals and leverage automated audio descriptors.
Another possible approach is computing image similarity descriptors and/or random projections on all the sections and represent a malware as \emph{bag of descriptors}, which can then be used for better characterization of malware.
Using the error model in the sparse representation based malware classification framework, we can determine the exact positions in which the malware variant differs from another variant.
This approach can also be used to find the exact source from which a malware variant evolves.
Patched malware that attaches to benign software can be identified using this method.

\section*{CONCLUSION}

In this paper we explored orthogonal yet complementary methods to analyze malware motivated by Signal and Image Processing.
Malware samples are represented as images or signals.
Image and signal based features are extracted to characterize malware.
Our extensive experiments demonstrate the efficacy of our methods on malware classification and retrieval.
We believe that our techniques will open the scope of signal and image based methods to broader fields in computer security.


\begin{thebibliography}{1}



\bibitem{gist2}
A.~Oliva and A.~Torralba, Modeling the shape of the scene: A holistic representation of the spatial envelope, International Journal of Computer Vision, 42(3), 145-175, 2002.

\bibitem{gist1}
A.~Torralba, K.P.~Murphy, W.T.~Freeman and M.~Rubin, Context-based vision system for place and object recognition, In Proceedings of the
Ninth IEEE International Conference on Computer Vision, pp. 273-280, 2003. 

\bibitem{gist-eval}
M.~Douze, H.~Jégou, H.~Sandhawalia, L.~Amsaleg and M.~Schmid, Evaluation of gist descriptors for web-scale image search, In Proceedings of the ACM International Conference on Image and Video Retrieval, (p. 19). , 2009. 

\bibitem{malw-imgs}
L.~Nataraj, S.~Karthikeyan, G.~Jacob and B.S.~Manjunath, Malware Images: Visualization and Automatic Classification, In Proceedings of the 8th International Symposium on Visualization for Cyber Security (VizSec '11), 2011, New York, NY, USA.


\bibitem{malgenome}
Y.~Zhou and X.~Jiang,  Dissecting android malware: Characterization and evolution, In Proceedings of the IEEE Symposium on Security and Privacy (SP), 2012.


\bibitem{malimg-ds}
Malimg Dataset, {\it http://old.vision.ece.ucsb.edu/spam/malimg.shtml}


\bibitem{malheur}
K.~Rieck, P.~Trinius, C.~Willems and T.~Holz, Automatic analysis of malware behavior using machine learning, Professoren der Fak. IV., 2009.


\bibitem{vxshare}
VirusShare, {\it http://www.virusshare.com}



\bibitem{comp-assm}
L.~Nataraj, V.~Yegneswaran, P.~Porras and J.~Zhang, A comparative assessment of malware classification using binary texture analysis and dynamic analysis, In In Proceedings of the 4th ACM Workshop on Security and Artificial Intelligence, (AISec '11), 2011, Chicago, IL, USA.

\bibitem{sigmal}
D.~Kirat, L.~Nataraj, G.~Vigna and B.S.~Manjunath, SigMal: A static signal processing based malware triage, In Proceedings of Proceedings of the 29th Annual Computer Security Applications Conference (ACSAC '13), 2013, New Orleans, LA, USA.


\bibitem{sarvam}
L.~Nataraj, D.~Kirat, B.S.~Manjunath and G.~Vigna, SARVAM: Search And RetrieVAl of Malware, In Proceedings of the Annual Computer Security Conference (ACSAC) Worshop on Next Generation Malware Attacks and Defense (NGMAD '13), 2013, New Orleans, LA, USA.


\bibitem{vtotal}
Virustotal, {\it https://www.virustotal.com/}

\bibitem{sattva}
L.~Nataraj, S.~Karthikeyan, and B.S.~Manjunath, SATTVA: SpArsiTy inspired classificaTion of malware VAriants, In Proceedings of the 3rd ACM Workshop on Information Hiding and Multimedia Security (IH\&MMSEC '15), 2015, Portland,OR, USA.




\bibitem{2009-pami-wright-face}
J.~Wright, A.Y.~Yang, A.~Ganesh, S.S.~Sastry and Y.~Ma, Robust face recognition via sparse representation, In IEEE Transactions on IPattern Analysis and Machine Intelligence, 31(2), 210-227, 2009. 

\bibitem{2009-donoho-counting}
D.~Donoho and J.~Tanner, Counting faces of randomly projected polytopes when the projection radically lowers dimension, In Journal of the American Mathematical Society, 22(1), 1-53, 2009. 


\bibitem{2011-jay-pami}
J.K.~Pillai, V.M.~Patel, R.~Chellapa and N.K.~Ratha, Secure and robust iris recognition using random projections and sparse representations, In IEEE Transactions on IPattern Analysis and Machine Intelligence, 33(9), 1877-1893, 2011. 


\end{thebibliography}
\end{document}